\newif\iflong
\longtrue
\iflong
  \documentclass[11pt,a4paper]{article}
  \usepackage{fullpage}
  \usepackage[english]{babel}
  \usepackage{amssymb}
  \usepackage[leqno]{amsmath}
  \makeatletter
  \newcommand{\leqnomode}{\tagsleft@true\let\veqno\@@leqno}
  \newcommand{\reqnomode}{\tagsleft@false\let\veqno\@@eqno}
  \makeatother

  \usepackage{mathabx}
  \usepackage{mathtools}
  \usepackage{csquotes}
  \usepackage[all]{nowidow}
  \usepackage{amsthm,amsmath,amssymb}
  \usepackage[mathcal,mathscr]{eucal}
  \usepackage{epsfig,graphicx,graphics,color}
  \usepackage{enumerate}
  \usepackage[sort&compress,numbers]{natbib}
\usepackage{hyperref}
\hypersetup{
    colorlinks=true,       % false: boxed links; true: colored links
    linkcolor=blue,        % color of internal links (change box color with linkbordercolor)
    citecolor=red,         % color of links to bibliography
    filecolor=magenta,     % color of file links
    urlcolor=cyan,         % color of external links
    linktocpage=true
}

\usepackage{bbm}
  
\theoremstyle{plain}
\newtheorem{theorem}{Theorem}
\newtheorem{lemma}[theorem]{Lemma}
\newtheorem{corollary}[theorem]{Corollary}

\newtheorem{proposition}[theorem]{Proposition}

\theoremstyle{definition}

\else
  \documentclass[runningheads]{llncs}
\fi

\usepackage{graphicx}
\usepackage{amsmath}
\usepackage{amssymb}
\usepackage{color}
\usepackage[]{algorithm2e}
\usepackage{tikz}
\usepackage{booktabs}
\usepackage{multirow}
\usepackage{hyperref}

\title{Single Machine Batch Scheduling to\\ Minimize the Weighted Number of Tardy Jobs}
\iflong
  \author{Danny Hermelin\thanks{Ben-Gurion University of the Negev, Beer-Sheva, Israel. \texttt{hermelin@bgu.ac.il}}
     \and Matthias Mnich\thanks{TU Hamburg, Institute for Algorithms and Complexity, Hamburg, Germany. \texttt{matthias.mnich@tuhh.de}}
     \and Simon Omlor\thanks{TU Hamburg, Institute for Algorithms and Complexity, Hamburg, Germany. \texttt{simon.omlor@tuhh.de}}}
 \date{}
\else
  \titlerunning{Single Machine Batch Scheduling}
  \author{Danny Hermelin\inst{1}
     \and Matthias Mnich\inst{2}
     \and Simon Omlor\inst{2}}
  \authorrunning{Hermelin, Mnich, Omlor}
  \institute{Ben-Gurion University of the Negev, Beer-Sheva, Israel. \email{hermelin@bgu.ac.il}
        \and TU Hamburg, Institute for Algorithms and Complexity, Hamburg, Germany. \email{\{matthias.mnich,simon.omlor\}@tuhh.de}}
\fi

\begin{document}
\maketitle              % typeset the header of the contribution

\begin{abstract}
  The $1|B,r_j|\sum w_jU_j$ scheduling problem takes as input a batch setup time $\Delta$ and a set of~$n$ jobs, each having a processing time, a release date, a weight, and a due date; the task is to find a sequence of batches that minimizes the weighted number of tardy jobs.
  This problem was introduced by Hochbaum and Landy in 1994; as a wide generalization of {\sc Knapsack}, it is $\mathsf{NP}$-hard.

  \iflong\else$\quad$\fi In this work we provide a multivariate complexity analysis of the $1|B,r_j|\sum w_jU_j$ problem with respect to several natural parameters.
  That is, we establish a thorough classification into fixed-parameter tractable and $\mathsf{W}[1]$-hard problems, for parameter combinations of\iflong\linebreak\fi (i) $\#p$ = distinct number of processing times, (ii) $\#w$ = number of distinct weights, (iii) $\#d$ = number of distinct due dates, (iv) $\#r$ = number of distinct release dates, and (v) $b$ = batch sizes.
  Thereby, we significantly extend the work of Hermelin et al. (2018) who analyzed the parameterized complexity of the non-batch variant of this problem without release dates.

  \iflong\else$\quad$\fi As one of our key results, we prove that $1|B,r_j|\sum w_jU_j$ is $\mathsf{W}[1]$-hard parameterized by the number of distinct processing times and distinct due dates.
  To the best of our knowledge, these are the first parameterized intractability results for scheduling problems with few distinct processing times.
  Further, we show that $1|B,r_j|\sum w_jU_j$ is fixed-parameter tractable with respect to parameter $\#p + \#d + \#r$ and with respect to parameter $\#w + \#d$ if there is just a single release date.
  Both results hold even if the number of jobs per batch is limited by some integer $b$.

  \iflong
  \else
  \fi
\end{abstract}

\section{Introduction}
\label{sec:introduction}
This paper is concerned with the problem of minimizing the total weight of tardy (late) jobs in a single machine batch scheduling environment. Before describing our results, we first briefly overview the classical non-batch variant of this problem, denoted as $1||\sum w_jU_j$ in Graham's classical three-field notation~\cite{GrahamEtAl79}.
Following this, we describe the extension of $1||\sum w_jU_j$ to the batch scheduling environment, and discuss how our results fit into the known state of the art. 

\iflong
  \subsection{Total weight of tardy jobs on a single machine}
\else
  \subsubsection*{Total weight of tardy jobs on a single machine.}
\fi
One of the most fundamental and prominent scheduling criteria on a single machine is that of minimizing the total weight of tardy jobs in a schedule.
Let~$J$ be a set of jobs, where each job $j \in J$ has a \emph{processing time} $p_j \in \mathbb{N}$, a \emph{weight} $w_j \in \mathbb{N}$, and a \emph{due date} $d_j \in \mathbb{N}$. %and a \emph{release time} $r_j\in \mathbb{N}$. 
We are given a single machine on which to process all the jobs in $J$.
A schedule for this machine corresponds to assigning a \emph{starting time} $S_j$ to each job $j \in J$, so that $S_i \notin [S_j,S_j+p_j)$ for any job $i \neq j$.
The term $S_j+p_j$, also denoted $C_j$, is called the \emph{completion time} of job~$j$.
A job $j\in J$ is \emph{tardy} if its completion time exceeds its deadline, i.e., if $C_j > d_j$; otherwise, it is \emph{early}.
The goal is to find a schedule which minimizes the total weight of all tardy jobs; or $\sum_{j \in J} w_jU_j$ where~$U_j$ is a binary indicator variable which takes value 1 if and only if job $j$ is tardy.
This problem is denoted as the $1||\sum w_jU_j$ problem.

Karp~\cite{Karp1972} proved that this problem is (weakly) $\mathsf{NP}$-hard even when all jobs have a common due date (i.e., the $1|d_j=d| \sum w_jU_j$ problem), and in fact this variant is equivalent to the \textsc{0/1 Knapsack} problem.
The variant where in addition to a single due date, the weight of each job is equal to its processing time (the $1|d_j=d,p_j=w_j| \sum w_jU_j$ problem) is known to be equivalent to the {\sc Partition} problem.  

Lawler and Moore~\cite{LawlerMoore1969} provided a pseudo-polynomial time algorithm for $1||\sum w_jU_j$, whereas Sahni~\cite{sahni1976} showed that the problem admits an FPTAS.
The variant where all jobs have unit weight (and a single release date), known as the $1||\sum U_j$ problem, is solvable in $O(n\log n)$ time due to an algorithm by Moore~\cite{Moore1968}.
There is also a classical variant where each job $j \in J$ also has a release time $r_j \in \mathbb{N}$, and $S_j \geq r_j$ is required of any schedule.
This variant is known to be $\mathsf{NP}$-hard even if jobs have unit weight and there are only two distinct due dates and only two distinct release times. 

%A fundamental problem in scheduling is to sequence a set $J$ of $n$ jobs, each of which is characterized by a processing time $p_j$ and a due date $d_j$, on a single machine, with the objective of minimizing the number of tardy jobs (jobs missing their due date). This problem was first considered by Moore~\cite{Moore1968}, who provided an $O(n\log n)$-time algorithm for finding an optimal schedule. In Graham's classical three-field notation, this problem is denoted as $1||\sum U_j$; the variable $U_j$ takes value 1 if job $j$ is tardy and value 0 otherwise.

%As soon as each job $j\in J$ additionally has a weight $w_j$, the scheduling problem $1||\sum w_jU_j$ becomes $\mathsf{NP}$-hard. This holds even if all jobs have the same due date; this fundamental intractability result was shown by Karp~\cite{Karp1972} in his landmark paper.

Most relevant to this paper is a recent result by Hermelin et al.~\cite{HermelinEtAl2018} who studied the $1||\sum w_jU_j$ problem from the perspective of parameterized complexity~\cite{CyganEtAl2015}.
There, the following three parameters are considered for the problem: 
\begin{itemize}
  \item $\#d$: number of distinct due dates,
  \item $\#p$: number of distinct processing times,
  \item $\#w$: number of distinct weights.
\end{itemize}
Their main results are given in the theorem below: 
\begin{proposition}[\cite{HermelinEtAl2018}]
  \label{thm:hermelin}
  Problem $1||\sum w_jU_j$ can be solved in
  \begin{itemize}
    \item time $f(\#d + \#p) \cdot n^{O(1)}$, time $f(\#d + \#w) \cdot n^{O(1)}$, and in time $f(\#p + \#w) \cdot n^{O(1)}$.
    \item time $n^{O(\#p)}$, and in time $n^{O(\#w)}$.
  \end{itemize}
\end{proposition}
A special case of this result was already obtained by Etscheid et al.~\cite{EtscheidEtAl2017} who presented an $f(\#p) \cdot n^{O(1)}$-time algorithm for the single due date $1|d_j=d|\sum w_jU_j$ problem.

%In that regard, the work of Hermelin et al.~\cite{HermelinEtAl2018} fits well into the growing area of designing parameterized algorithms for scheduling problems~\cite{HermelinEtAl2019,KnopKoutecky2018,KnopEtAl2017,MnichWiese2015,vanBevernEtAl2016,vanBevernEtAl2015,vanBevernEtAl2017}. For background, we refer to the recent survey on the topic~\cite{MnichvanBevern2018}.
%For the special case of $\#d = 1$ which was shown to be $\mathsf{NP}$-hard by Karp~\cite{Karp1972}

\iflong
  \subsection{Batch scheduling}
\else
  \subsubsection*{Batch scheduling.}
\fi
Batch scheduling has recently received a considerable amount of attention in the scheduling community.
The motivation for this line of research stems from the fact that in manufacturing systems items flow between facilities in boxes, pallets, or carts.
A set of items assigned to the same container is considered as a \emph{batch}.
It is often the case that items in the same batch leave the facility together, and thus have equal completion time.
We refer to Potts and Kovalyov~\cite{PottsKovalyov2000} and Webster and Baker~\cite{WebsterBaker1995} for further reading on the topic.

Hochbaum and Landy~\cite{HochbaumLandy1994} studied the generalization of the $1||\sum w_jU_j$ problem to the batch setting.
In this problem, denoted $1|B|\sum w_jU_j$, a schedule consists of a partition of the job set $J$ into batches, and a starting time $S_B$ for each batch $B$ such that $S_{B'} \notin [S_B, C_B = S_B + \Delta + \sum_{j \in B} p_j)$ for any batch $B' \neq B$, where $\Delta$ is a given \emph{setup time} associated with starting any batch.
The completion time of any job $j \in B$ is $C_j = C_B$, meaning that all the jobs together in a batch are completed at the same time. The goal is again to minimize the total weight of tardy jobs $\sum w_jU_j$.
Note that the order of the jobs within each batch is irrelevant, and that when $\Delta = 0$ this problem becomes the classical $1||\sum w_jU_j$ problem. 

%A robust generalization of $1||\sum w_jU_j$ was introduced by Hochbaum and Landy. In their seminal work~\cite{HochbaumLandy1994}, they consider the {\sc Weighted Tardiness with Batching} problem: given a batch setup time $\Delta$ and set $J$ of $n$ jobs each characterized by a processing time $p_j$, a due date $d_j$ and a weight $w_j$, the objective is to schedule the jobs in batches so as to minimize the weighted number of tardy jobs. This problem is denoted by $1|B|\sum w_jU_j$. The key feature of $1|B|\sum w_jU_j$ is starting a new batch incurs a setup cost of $\Delta$, while the completion time of each job is equal to the completion time of its batch, so that all the jobs together in a batch are completed at the same time.
Hochbaum and Landy observed that this problem is weakly $\mathsf{NP}$-hard (being a direct generalization of $1||\sum w_jU_j$), and provided pseudopolynomial-time algorithms for the problem that are linear in the total sum of job processing-times (plus $n\cdot \Delta$) or the maximum due-date.
Brucker and Kovalyov provided an analogous algorithm which is linear the total sum of job weights~\cite{BruckerKoyalyov1996}.
Nevertheless, in this paper we are interested in the case where job weights, processing times, or due dates can be arbitrarily large, but the number of different values of each of these parameters (namely, $\#w$, $\#p$, or $\#d$) is relatively small.
In this context, the following result of Hochbaum and Landy is very relevant.
\begin{proposition}[\cite{HochbaumLandy1994}]
\label{thm:Hochbaum}%
  Problems $1|B,p_j=p|\sum w_jU_j$ and $1|B|\sum U_j$ are polynomial-time solvable. 
\end{proposition}

One can also consider restrictions on batches that are relevant in practice.
For instance, one can require a bound on the \emph{size} $|B|$ or  \emph{volume} $||B||=\sum_{j\in B}p_j$ of any batch $B$.
Cheng and Kovalyov~\cite{ChengKovalyov2001} argued about the importance of the batch-size $|B|\leq b$ bound in real-life applications.
Note that for $b = n$ we have the unbounded $1|B|\sum w_jU_j$ problem, whereas for $b = 1$ one obtains the classical non-batch model $1||\sum w_jU_j$.
The following is a very relevant result of Cheng and Kovalyov who showed that $1||B|\leq b|\sum U_j$ is in XP when parameterized by either $\#p$ or $\#d$:
\begin{proposition}[\cite{ChengKovalyov2001}]\label{thm:ChengKovalyov}
  Problem $1||B|\leq b|\sum U_j$ can be solved in time $n^{O(\#p)}$, and in time $n^{O(\#d)}$.
\end{proposition}

\iflong
  \subsection{Our contributions}
\else
  \subsubsection*{Our contributions.}
\fi
\label{sec:ourcontributions}
We provide a thorough multivariate complexity analysis of $1|B|\sum w_jU_j$ and related variants: Problem $1|B,r_j|\sum w_jU_j$ where jobs also have release dates, problem $1|\,|B|\leq b|\sum w_jU_j$ where there is a bound on the batch size, and problem $1|\,||B||\leq b|\sum w_jU_j$ where there is a bound on the batch volume.
%A summary of our results is given in Table~\ref{tab:results}.

\medskip
\noindent \emph{The standard batch model:} In the first part of the paper we study the $1|B|\sum w_jU_j$ problem without release dates or batch restrictions.
We show that almost all results of Proposition~\ref{thm:hermelin} regarding the $1|| \sum w_jU_j$ problem extend to the batch setting. 
\begin{theorem}
\label{thm:standard}%
  Problem $1|B|\sum w_jU_j$ can be solved in
  \begin{itemize}
    \item time $n^{O(\#p)}$, and in time $n^{O(\#w)}$.
    \item time $f(\#d + \#p) \cdot n^{O(1)}$, and in time $f(\#d + \#w) \cdot n^{O(1)}$.
  \end{itemize}
\end{theorem}
The second part of this theorem is proved by an elegant reduction to the non-batch case, while the first part is based on dynamic programming.
Note that the second item of the theorem is a generalization of the result by Hochbaum and Landy stated in Proposition~\ref{thm:Hochbaum}.
\medskip

%Recall that Karp's intractability result for $1|d_j=d|\sum w_jU_j$ implies that $1|B|\sum w_jU_j$ is NP-hard for parameter $\#d = 1$. Thus, the only remaining case that 

\noindent \emph{Release dates:} Next, we show that adding release dates makes the problem much harder.
%Precisely, we allow each job $j\in J$ to have its individual release date $r_j$, only after which it can be sequenced on the machine.
%Previous works thus considered the setting $r_j = 0$ for all $j\in J$, whereas we take a look at $\#r$, the number of distinct release dates.
%Here, Karp's intractability result implies that $1|B|\sum w_jU_j$ is $\mathsf{NP}$-hard for parameter $\#d = \#r = 1$.
Specifically, we prove that $1|B,r_j|\sum w_jU_j$ is highly unlikely to be fixed-parameter tractable for parameter $\#d+\#p$ or $\#p+\#r$.
\begin{theorem}
\label{thm:release}%
  Problem $1|B, r_j|\sum w_jU_j$ is $\mathsf{W}[1]$-hard when parameterized by $\#d + \#p$, and is $\mathsf{W}[1]$-hard when parameterized by $\#p + \#r$.
  Furthermore, the problem is solvable in time\iflong\linebreak\fi $n^{f(\#p + \#r, + \#w)}$, and in time $n^{f(\#p + \#d + \#w)}$.
\end{theorem}

\noindent To the best of our knowledge, this is the first $\mathsf{W}[1]$-hardness result for any scheduling problem parameterized by the number of distinct processing times $\#p$.
In particular, whether or not $P||C_{\max}$ (makespan minimization on an unbounded number of parallel machines) is $\mathsf{W}[1]$-hard for this parameter is a famous open problem (see~\cite{MnichvanBevern2018}), and this question is also open for $1||\sum w_jU_j$~\cite{HermelinEtAl2018}.
\medskip

%\noindent Further we show that we can solve Problem $1|B, r_j|\sum w_jU_j$ in time $n^f(\#d, \#p, \#w)$ and in time $n^f(\#d, \#p, \#w)$.

%For the only remaining two-dimensional parameter, $\#p + \#w$, we show the following:

%Those intractability results justify consideration of three-dimensional parameters.

%In contrast, our third main result is a fixed-parameter algorithm for $1|B, r_j|\sum w_jU_j$ parameterized by $\#d + \#p + \#r$.
%\begin{theorem}
%\label{thm:weighted-tardy-batches-reltimes-dpr}
%Problem $1|B,r_j|\sum w_jU_j$ is fixed-parameter tractable for parameter $\#d+\#p+\#r$.
%\end{theorem}
%For parameter $\#d + \#r + \#w$, ...

\noindent \emph{Batch restrictions:} In the final part of the paper we show that the $f(\#d + \#w) \cdot n^{O(1)}$ algorithm in the second part of Theorem~\ref{thm:standard} can be generalized to the setting where each batch contains at most $b$ jobs; this setting was proposed by Cheng and Kovalyov~\cite{ChengKovalyov2001}.
Further, the algorithm with run time $f(\#d + \#p) \cdot n^{O(1)}$ can be generalized to the setting where the batch size is limited and the jobs may have different release dates.
%Finally, we study the model with bounded batch sizes, which was proposed by Cheng and Kovalyov~\cite{ChengKovalyov2001}.
%In this model we propose the following fixed-parameter algorithm
\begin{theorem}
\label{thm:weighted-tardy-boundedbatchsize-dw}%
  The following problems are fixed-parameter tractable:
  \begin{itemize}
    \item $1||B| \leq b|\sum w_jU_j$ for parameter~$\#d + \#w$.
    \item $1||B| \leq b,r_j|\sum w_jU_j$ for parameter~$\#d + \#p + \#r$.
  \end{itemize}
\end{theorem}
In particular, this improves the result of Cheng and Kovalyov stated in Proposition~\ref{thm:ChengKovalyov}, as our algorithm runs in time $f(\#d)\cdot n^{O(1)}$ for the unweighted version $1||B| \leq b|\sum U_j$.

Finally, let us make a few remarks on the problem $1|||B|| \leq V|\sum w_jU_j$, where the batch volume is bounded.
First, for this problem we show $\mathsf{NP}$-hardness even for the case of unit weights and a single due date; this rules out the existence of $\mathsf{XP}$-algorithms parameterized by $\#p+\#w$.
Second, we show that for parameter $\#p$, this problem is at least as hard as $P||C_{\max}$ parameterized by $\#p$.
Recall that the fixed-parameter tractability of $P||C_{\max}$ parameterized by~$\#p$ is a long-standing open problem.

%Recall that the problem $P||C_{\max}$ of scheduling a set of $n$ jobs on $m$ parallel machines so as to complete all jobs by some target time $T$ can be solved in time $(\log \Delta)^{2^{O(\#p)}}$, by the breakthrough result of Goemans and Rothvoss~\cite{GoemansRothvoss2014}.
%Resolving the parameterized complexity status of $P||C_{\max}$ for parameter $\#p$ is one of the 15 open problems in the recent survey~\cite{MnichvanBevern2018}.
%We show that resolving the parameterized complexity of $1|B,||B|| \leq V|\sum U_j$ for parameter $\#p + \#d$ is at least as hard as resolving that open problem.
%\begin{theorem}
%\label{thm:batch-volumepd}%
%Any instance $\mathcal I$ of $P||C_{\max}$ with $\#p$ different processing times can be transformed to an instance of $1|B,||B|| \leq V|\sum U_j$ with $\#p$ different processing times and a single due date, such that all jobs of $\mathcal I$ complete by time $T$ if and only if all jobs of $\mathcal I'$ are early.
%\end{theorem}

A summary of our results is given in Table~\ref{tab:results}.
\begin{table}[h]
  \centering
  \begin{tabular}{llll}
    \toprule
    Problem variant       & Parameters  & Result & Reference\\
    \midrule
    $1|B|\sum w_jU_j$     & $\#d$       & para-$\mathsf{NP}$-hard~~~ & Karp~\cite{Karp1972}\\
                          & $\#p$       & $\mathsf{XP}$ & Theorem~\ref{thm:standard}
                          \\
                          & $\#w$       & $\mathsf{XP}$ & Theorem~\ref{thm:standard}
                          \\
                          & $\#d+\#p$       & $\mathsf{FPT}$ & Theorem~\ref{thm:standard}
                          \\
                          & $\#d+\#w$       & $\mathsf{FPT}$ & Theorem~\ref{thm:standard}
                          \\
                          & $\#p + \#w$ & ?\\
    \midrule
    $1|B,|B|\leq b|\sum w_jU_j$ & $\#d + \#w$ & $\mathsf{FPT}$ & Theorem~\ref{thm:weighted-tardy-boundedbatchsize-dw}\\
    \midrule
    %\cmidrule{2-4}
    $1|B,r_j|\sum w_jU_j$ & $\#p + \#r$                  & $\mathsf{W}[1]$-hard~~~ & Theorem~\ref{thm:release}\\
                          & $\#d + \#p$ & $\mathsf{W}[1]$-hard & Theorem~\ref{thm:release}\\
    %\cmidrule{2-4}
                          & $\#p + \#w$                  & ?\\
    %\cmidrule{2-4}

                          & $\#d + \#r+\#w$ & para-$\mathsf{NP}$-hard~~~& Karp~\cite{Karp1972}\\
    %\cmidrule{2-4}
                          & $\#d + \#p + \#w$            & $\mathsf{XP}$ & Theorem~\ref{thm:release}\\
    %\cmidrule{2-4}
                          & $\#p + \#r + \#w$            & $\mathsf{XP}$ & Theorem~\ref{thm:release}\\
    \midrule
    $1|B,|B|\leq b, r_j|\sum U_j$~~~ & $\#d + \#p + \#r$~~~ & $\mathsf{FPT}$ & Theorem~\ref{thm:weighted-tardy-boundedbatchsize-dw} \\
    \bottomrule\\
  \end{tabular}
  \caption{Summary of results.\label{tab:results}}
\end{table}

\iflong
\else
  Due to space constraints, proofs of statements are deferred to the full version of the paper, which is available at \url{tbd}.
\fi

\section{The standard batch model}
\label{sec:basic_problem}
\begin{figure}[!ht]
  \begin{center}
    \includegraphics[scale=0.5]{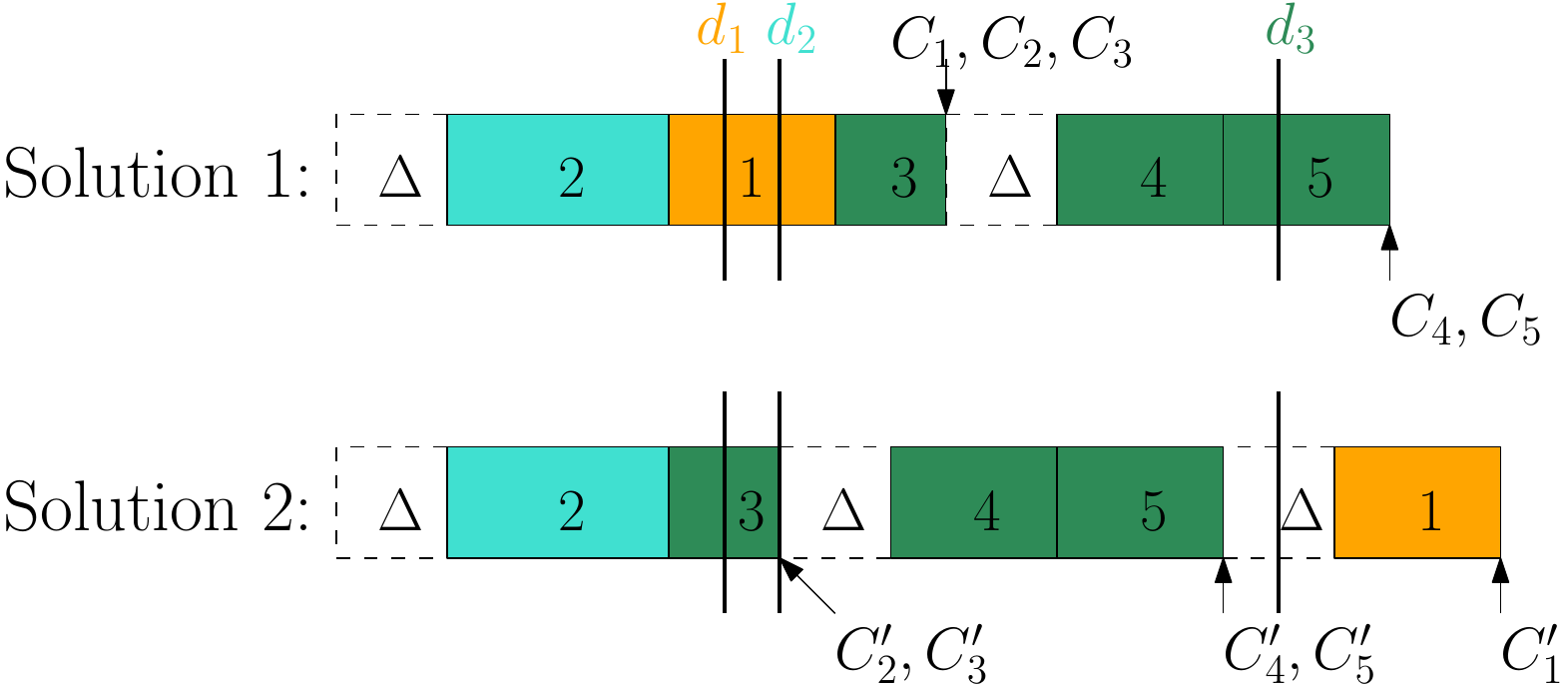}
  \end{center}
  \caption{An example for batch scheduling with 5 jobs.
    In the first solution job $1$ and 5 are tardy.
    In the second solution only job 1 is tardy.
    Tardy jobs can be moved to the end of the schedule without increasing the weight of the tardy jobs. \label{fig:batchscheduling}}
\end{figure}

In this section we present algorithms for the basic $1|B|\sum w_jU_j$ problem, providing a complete proof for Theorem~\ref{thm:standard}.
The proof is split into two parts, which are proven in three separate lemmas below.
Note that in the setting where all jobs are released at the same time and the batch sizes are not restricted we can schedule the early jobs in order of the due dates.
This is a very helpful observation by Hochbaum and Landy~\cite{HochbaumLandy1994}, which will be used multiple times in this section.
We illustrate it by an example in Fig.~\ref{fig:batchscheduling}.

\begin{lemma}[\cite{HochbaumLandy1994}]
\label{lem:EDD}%
  Any instance of $1|B|\sum w_jU_j$ admits an optimal solution in which all early jobs are in earliest due date (EDD) order.
  That is, for any two jobs $i$ and $j$ scheduled in two different batches $i \in B_1$ and $j \in B_2$ with $S_{B_1} < S_{B_2}$, we have $d_i < d_j$.
\end{lemma}

We use following notation to order the due dates: $d^{(1)}<d^{(2)} \dots < d^{(\#d)}$.
Further, we set~$d^{(0)}$ to be the smallest release date.

\subsection{Fixed-Parameter Algorithms}
We begin by presenting fixed-parameter algorithms for $1|B|\sum w_jU_j$ for parameter $\#d + \#p$, and for \mbox{$\#d + \#w$}.
\begin{lemma}
  Problem $1|B|\sum w_jU_j$ is solvable in time $f(\#d + \#p) \cdot n^{O(1)}$, and in time $f(\#d + \#w) \cdot\iflong\linebreak\fi n^{O(1)}$.
\end{lemma}
\iflong
\begin{proof}
  Let $J$ denote the job set of our $1|B|\sum w_jU_j$ instance.
  We first observe that there is an optimal schedule in which at most one batch completes within each interval $(d^{(i-1)}, d^{(i)}]$, for each $i \in \{1,\ldots,\#d\}$; if there are two or more batches ending in $(d^{(i-1)}, d^{(i)}]$, then these batches can be combined into a single batch without creating new tardy jobs.
  The second observation is that if there is no batch ending in $(d^{(i-1)}, d^{(i)}]$ then all jobs with due date $d^{(i)}$ that are completed early must be in batches ending at $d^{(i-1)}$ or earlier.
  We next use these observations to reduce our $1|B|\sum w_jU_j$ instance $J$ into $2^{O(\#d)}$ instances of the non-batch $1||\sum w_jU_j$ problem, each with the same number of processing times, weights, and due dates as in $J$.
  Combined with the fixed-parameter algorithms for $1||\sum w_jU_j$ given by Hermelin et al.~\cite{HermelinEtAl2018}, this will provide a proof for the theorem. 
 
  For each $i \in \{1,\ldots,\#d\}$, we guess whether there is a batch ending in $(d^{(i-1)}, d^{(i)}]$ in an optimal solution.
  Let $I\subseteq \{1,\dots,\#d\}$ be the set of indexes $i$ such that there is a batch ending in $(d^{(i-1)}, d^{(i)}]$ with respect to our guess.
  For an index $\ell \in \{1,\ldots,\#d\}$, let $I_{\leq \ell} =\{i \in I\mid i \leq \ell\}$ denote the set of indices in $I$ smaller or equally to $\ell$ and let $i(\ell)=\max\{i \in I ~|~ i\leq \ell\}$ be the largest index in $I$ that is less or equal than $\ell$. 
  We construct an instance $J_I$ of $1||\sum w_jU_j$ corresponding to $I$ by replacing the due date $d_j= d^{(\ell)}$ of each job $j\in J$ with an alternative due date $d'_j=d^{(i(\ell))}-|I_{\leq \ell}|\cdot \Delta$; all other job parameters remain the same in $J_I$.

  Consider some set of indices $I\subseteq \{1,\dots,\#d\}$, and let $J_I$ be the corresponding $1||\sum w_jU_j$ instance.
  We can convert a schedule of $J_I$ as follows:
  We note that
  \begin{equation*}
    \sum_{j\text{ is early and $d_j'\leq d$}}p_{j} \leq d
  \end{equation*}
  for all due dates $d$.
  We construct $|I|+1$ batches.
  The first $|I|$ batches are denoted by $B_i$ for $i\in I$ and are processed in increasing order, i.e. if $i < i'$ then $B_i$ is processed before $B_{i'}$.
  Let $j$ be an early job (i.e. $j'$ is early) with $d_j=d^{(\ell)}$ for some $\ell$.
  Then we assign $j$ to batch $B_{i(\ell)}$.
  We conclude that $j$ will be early as the completion time of $B_{i(\ell)}$ is equal to 
  \begin{equation*}
    C_{B_{i(\ell)}}=|I_{\leq \ell}|\Delta + \sum_{j'\text{ is early and $d_{j'}\leq d_j$}}p_{j'} \leq |I_{\leq \ell}| \Delta+d_j' =d^{i(\ell)} \leq d_j \enspace .
  \end{equation*}
  Conversely, consider any schedule for $J$ that schedules at most one batch ending in each interval of consecutive due dates, and let $I\subseteq \{1,\dots,\#d\}$ be the corresponding set of indices.
  Then any early job $j\in J$ with $d_j=d^{(l)}$ has $C_j \leq d^{i(\ell)}$, and so its completion time in the non-batch setting under the same ordering of early jobs is at most $C_j - |I(\leq \ell)| \Delta \leq d^{i(\ell)} - |I(\leq\ell)|\Delta = d'_j$. 
  
  It follows that an optimal schedule for our original $1|B|\sum w_jU_j$ instance corresponds to the schedule with the minimum weight of tardy jobs among all optimal schedules for instances $J_I$, $I\subseteq \{1,\dots,\#d\}$.
  The lemma then follows since there are $2^{\#d}$ instances $J_I$, and each instance can be solved in $f(\#d + \#p) \cdot n^{O(1)}$ or $f(\#d + \#w) \cdot n^{O(1)}$ time using the algorithm by Hermelin et al.~\cite{HermelinEtAl2018}.
\end{proof}
\fi

\subsection{XP algorithms} 
Assume that our input job set $\{1,\ldots,n\}$ is ordered such that $d_1 \leq \cdots \leq d_n$ (i.e. ordered according to EDD). 
Due to Lemma~\ref{lem:EDD}, there is an optimal schedule where any job $j \in J$ is either late, or it is scheduled after the early jobs in $\{1,\ldots,j-1\}$.
Thus, an optimal schedule for jobs $\{1,\ldots,j\}$ can be found by appending $j$ to some schedule of jobs $\{1,\ldots,j-1\}$.
As observed by Hochbaum and Landy~\cite[add cite]{HochbaumLandy1994}, when appending $j$ to such a schedule, there are three possibilities:
\begin{itemize}
  \item[a.] Job $j$ is included in the last batch of early jobs. 
  \item[b.] Job $j$ is included a new batch by itself, scheduled right after the previous last batch.
  \item[c.] Job $j$ is tardy. 
\end{itemize}
Below we devise two dynamic programming algorithms that utilize this fact. 
\begin{lemma}
\label{lem:p}%
  Problem $1|B|\sum w_jU_j$ is solvable in time $n^{O(\#p)}$.
\end{lemma}
\iflong
\begin{proof}
  Let $J= \{1,\ldots,n\}$ denote our job set ordered according to EDD, and let $p^{(1)} < \cdots < p^{(\#p)}$ denote the different processing times of all jobs in $J$.
  For increasing values of $j \in \{1,\ldots,n\}$, we compute a table~$W_j$ which has $n^{O(\#p)}$ entries and corresponds to jobs in $\{1,\ldots,j\}$. 

  The table $W_j$ will be indexed by a $\#p$-dimensional vector $I\in\{1,\ldots,n\}^{\#p}$, and integer $b \in \{1,\ldots,n\}$, and a due date $d \in \{0,d_1,\ldots,d_n\}$.
  The invariant that our algorithm will maintain is that $W_j[I,b,d]$ will equal the minimum total weight of tardy jobs in a schedule for jobs $\{1,\ldots,j\}$ with the following properties:
  \begin{enumerate}
    \item The early jobs are scheduled in EDD fashion as in Lemma~\ref{lem:EDD}.
    \item There are exactly $b$ batches containing exactly $I[i]$ early jobs, $i \in \{1,\ldots,\#p\}$, with processing time $p^{(i)}$, scheduled consecutively starting from time 0.
    \item The earliest due date among all jobs in the last batch is at least $d$.
  \end{enumerate}

  Note that there exists vector $I$ and integers $b$ and $d$ such that the optimal schedule for $J$ satisfies all properties of required from a schedule corresponding to entry $W_n[I,b,d]$ and all jobs in the first $b$ batches are early.

  In the beginning, we set $W_j[I,b,d]=\sum_{i=1}^j w_i$ if $I= \emptyset$, and $W_j[I,b,d]=\infty$ otherwise.
  Fix $j \in \{1,\ldots,n\}$, and consider an entry $W_j[I,b,d]$ of $W_j$. 
  Let $p^{(\ell)}=p_j$ be the processing time of~$j$ for $\ell \in \{1,\ldots,\#p\}$.
  Let~$I_\ell$ be the vector which coincides with $I$ on every coordinate, except for the $\ell^{\textnormal{th}}$ coordinate for which it is equal to $I[\ell]-1$.
  If the $\ell^{\textnormal{th}}$ coordinate of $I$ is $0$, then we set $W_j[I,b,d]=W_{j-1}[I,b,d]+w_j$.
  
  Now we consider the expression $\sum_{i=1}^{\#p} I[i] \cdot p^{(i)} + b\Delta$.
  If $\sum_{i=1}^{\#p} I[i] \cdot p^{(i)} + b\Delta > d$, then job $j$ will be late if it is among the jobs scheduled in the first $b$ batches.
  Since all of the first $j$ jobs with processing time $p_j$ have a due date less or equal to $d_j$, there cannot be a schedule that schedules exactly~$I[i]$ early jobs with processing time $p^{(i)}$ if we consider only the first $j$ jobs.
  Thus, we set $W_j[I,b,d]=\infty$.
  
  Else, if $\sum_{i=1}^{\#p} I[i] \cdot p^{(i)} + b\Delta \leq d$, we can schedule job $j$ early.
  There are two possibilities to do so.

  The first possibility is to schedule job $j$ in an already existing batch.
  Then the total weight of tardy jobs is $W_{j-1}[I_\ell,b,d]$.
  
  The second possibility is to open a new batch for job $j$.
  Then we look at the entries $W_{j-1}[I_\ell,b-1,d'] $ for $d' \leq d$.
  
  There is also the possibility to schedule $j$ tardy.
  In this case, the weight is given by $~W_{j-1}[I,b,d] + w_j $.
  Then the recursion for $W_j[I,b,d]$ is given by 
  \begin{equation*}
    W_j[I,b,d]= \min \left\lbrace W_{j-1}[I_\ell,b,d],~\min_{d'\leq d}\{W_{j-1}[I_\ell,b-1,d'] \},~W_{j-1}[I,b,d] + w_j \\
\right\rbrace \enspace .
  \end{equation*}

  %Observe that the three cases correspond to the three options of adding $j$ to a schedule of $\{1,\ldots,j-1\}$ discussed above.
  %In the first case we add $j$ to the last batch of early jobs.
  %The second option is to place $j$ in a new batch on its own. 
  %Finally, the third option corresponds to scheduling job $j$ as tardy. 

  Correctness of our dynamic programming algorithm is immediate following the discussion above.
  The optimal schedule corresponds to the minimum entry $W_n[I,b,d]$ over all $I\in\{1,\ldots,n\}^{\#p}$, $b \in \{1,\ldots,n\}$, and $d \in \{0,d_1,\ldots,d_n\}$.
  Note that since table $W_j$ has $n^{O(\#p)}$ entries, and each entry requires $O(1)$ time, computing the entire table can be done in $n^{O(\#p)}$.
  Thus, the algorithm for computing all tables $W_j$ has the same running time, and the lemma follows.
\end{proof}
\fi

\begin{lemma}
  Problem $1|B|\sum w_jU_j$ is solvable in time $n^{O(\#w)}$. 
\end{lemma}
\iflong
\begin{proof}
  Let $J=\{1,\ldots,n\}$ denote our job set ordered according to EDD, and let $w^{(1)} < \cdots < w^{(\#w)}$ denote the different weights of all jobs in $J$.
  The algorithm is very similar to the algorithm in the proof of Lemma~\ref{lem:p}, except here we compute tables $P_j$ that store minimum total processing time of early jobs, as opposed to minimum total weight of tardy jobs.
  Namely, for $I\in\{1,\ldots,n\}^{\#p}$, $b \in \{1,\ldots,n\}$, and $d \in \{0,d_1,\ldots,d_n\}$, entry $P_j[I,b,d]$ will equal the minimum total processing time of the early jobs in a schedule for jobs $\{1,\ldots,j\}$ that satisfies the all properties required in the proof of Lemma~\ref{lem:p}, except that the second condition is rephrased to require exactly $I[i]$ early jobs, $i \in \{1,\ldots,\#w\}$, with weight $w^{(i)}$.

  Fix $j \in \{1,\ldots,n\}$, and let $\ell \in \{1,\ldots,n\}$ denote the index such that $w_j =w^{(\ell)}$.
  The base cases for computing $P_j[I,b,d]$ are very similar to those described in the proof of Lemma~\ref{lem:p}:
  
  If $ P_{j-1}[I_\ell,b,d] + p_j > d$ and $\min_{d' \leq d} \{ P_{j-1}[I_\ell,b-1,d'] + p_j + \Delta \} > d $ or if $d>d_j$ then we cannot schedule exactly $I[i]$ jobs with weight $w^{(i)}$ early \emph{including} job $j$ if we consider only the first $j$ jobs.
  Thus, we set $P_j[I,b,d]=P_{j-1}[I,b,d] $.
  
  Otherwise, the main recursive formula is given by
  \begin{equation*}
    P_j[I,b,d]= \min \left\lbrace P_{j-1}[I_\ell,b,d] +p_j,~\min_{d' \leq d} \{ P_{j-1}[I_\ell,b-1,d'] + p_j + \Delta \},~P_{j-1}[I,b,d] 
\right\rbrace \enspace .\qedhere
  \end{equation*}
\end{proof}
\fi

%%%%%%%%%%%%%%%%%%%%%%%%%%%%%%%%%%%%%%%%%%%%%%%
%%%%% The standard batch model
%%%%%%%%%%%%%%%%%%%%%%%%%%%%%%%%%%%%%%%%%%%%%%%
\section{Release dates}
\label{sec:parameterizedintractabilityresults}
In this section we show that the problem of minimizing the weighted number of tardy jobs on a single batch machine when release dates are present is $\mathsf{W}[1]$-hard for parameters $\#p+\#r$ and $\#p+\#d$.
That is, we prove Theorem~\ref{thm:release}.
Thereafter, we give $\mathsf{XP}$-algorithms for $1|B,r_j|\sum w_jU_j$ parameterized by $\#p+\#w+\#r$, and parameterized by $\#p+\#w+\#d$.

We begin with parameter $\#p+\#r$; the hardness for parameter $\#p+\#d$ will follow almost immediately afterwards. 
To prove that $1|B,r_j|\sum w_jU_j$ is $\mathsf{W}[1]$-hard with respect to \mbox{$\#p+\#r$}, we present a reduction from the {\sc $k$-Sum} problem.
In this problem, we are given a set $\{x_1,\ldots,x_n\}$ of $n$ positive integers, and a target integer~$t$.
The task is to decide if there exist $k$ (not necessarily distinct) integers $x_{\pi(1)},\ldots, x_{\pi(k)} \in \{x_1,\ldots,x_n\}$ that sum up to $t$.
Abboud, Lewi, and Williams~\cite{AbboudEtAl2014} showed that {\sc $k$-Sum} is $\mathsf{W[1]}$-hard parameterized by $k$, even if all integers are in the range $\{1, 2, \ldots, n^{ck}\}$ for some constant $c$.

\subsection{The construction}
Let $(x_1,\ldots,x_n;t)$ be an instance of {\sc $k$-Sum}, with $x_i \in \{1, 2, \ldots, n^{ck}\}$ for each $i$.
Observe that due to their small range, each input integer $x_i$ can be written in the form $x_i = \sum_{j=0}^{ck} \alpha_{i,j} \cdot n^j$ for integers $\alpha_{i,0},\ldots,\alpha_{c,k} \in \{0,\ldots,n-1\}$, i.e., the \emph{base $n$ representation} of $x_i$.
We will heavily exploit this property in our construction.

Write $X = \sum_i x_i$.
Furthermore, we will assume throughout that $k-1$ times the largest integer in $\{x_1,\ldots,x_n\}$ is less than $t$.
If this is not the case, one can slightly modify the input by adding $kn^{ck}$ to each integer, and setting the target to $t+k^2n^{ck}$.
We construct an instance of $1|B, r_j|\sum w_jU_j$ with $O(k)$ distinct processing times and release times, such that there exists a feasible schedule with $\sum_j w_jU_j \leq k X-t +(n-1)k$ to if and only if there exist $k$ integers $x_{\pi(1)},\ldots, x_{\pi(k)} \in \{x_1,\ldots,x_n\}$ that sum up to $t$:
\begin{itemize}
  \item We create $(k-1)t$ identical jobs, referred to as \emph{leftover jobs}, each with the following parameters:
  \begin{itemize}
    \item Processing time 1 and weight $k(X+n)$.
    \item Release time 0 and due date $3kt$. 
  \end{itemize}
  \item For each $\ell \in \{1,\ldots,k\}$, and each input integer $x_i = \sum_{j=0}^{ck} \alpha_{i,j} \cdot n^j$, we create a set $J_{i,\ell}$ of \emph{normal jobs} that corresponds to $x_i$.
    This set consists of $\alpha_{i,j}$ jobs, for each $j \in \{0,\ldots,ck\}$, with the following parameters:
    \begin{itemize}
      \item Processing time $n^j$ and weight $n^j+n^j/x_i$.
      \item Release time $r_\ell = (\ell-1)3t$ and due date $(\ell-1)3t + t + x_i$. 
    \end{itemize}
  \item The batch setup time is set to $\Delta=t$.
  \item The bound on the total weight of tardy jobs is set to $k X-t +(n-1)k$.
\end{itemize}

Observe that the total processing time of all jobs in the set $J_{i,\ell}$ is precisely~$x_i$, and their total weight is~$x_i +1$.
This will be crucial later on.
Also note that whereas the weights above are fractional, one can make them integral by multiplying with $\prod x_i$. 

\subsection{Correctness}
\begin{figure}[h]
  \begin{center}
     \includegraphics[scale=0.8]{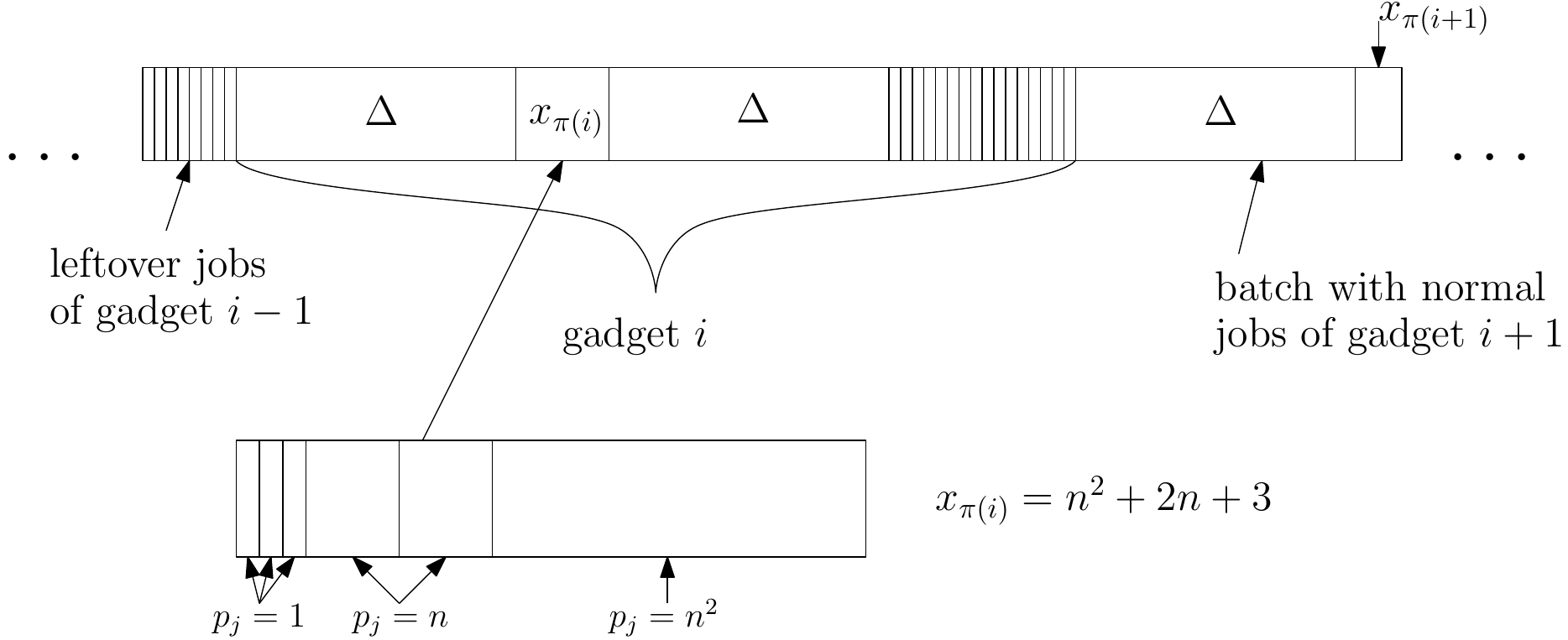}
  \end{center}
  \caption{An illustration of what the schedule given in Lemma~\ref{lem:EasyDirection} looks like. \label{fig:W1hardness}}
\end{figure}

\begin{lemma}
\label{lem:EasyDirection}%
  Suppose there exist $x_{\pi(1)},\ldots, x_{\pi(k)} \in \{x_1,\ldots,x_n\}$ such that $\sum_i x_{\pi(i)} = t$.
  Then there exists a schedule with $\sum_j w_jU_j \leq k X-t +(n-1)k$.
\end{lemma}
\iflong
\begin{proof}
  We create a schedule with $2k+1$ batches $B_1,\ldots,B_{2k+1}$.
  For $\ell \in \{1,\ldots,k\}$, we schedule all jobs in the set $J^{(\ell)}_{\pi(\ell)}$ in batch $B_{2\ell-1}$, and $t-x_{\pi(\ell)}$ leftover jobs in batch $B_{2\ell}$.
  We schedule the starting time of batch $B_{2\ell-1}$ at time $3t(\ell-1)$, and batch $B_{2\ell}$ at time $3t(\ell-1)+t+x_{\pi(\ell)}$.
  The remaining jobs are all scheduled in batch $B_{2k+1}$ which starts at time $3kt$.
  Note that in this way all jobs are scheduled after their release times, and only jobs in the last batch $B_{2k+1}$ are tardy.
  An easy calculation shows that the total weight of jobs in this last batch is 
  \begin{equation*}
    \sum_{j\in B_{2k+1}} w_j = kX+kn-\sum_{i=1}^k (x_{\pi(i)}+1)
                             = kX-K+(n-1)k \enspace .\qedhere
  \end{equation*}
  %The lemma thus follows.
\end{proof}
\fi
We illustrate Lemma~\ref{lem:EasyDirection} by an example in Fig.~\ref{fig:W1hardness}.

The converse of Lemma~\ref{lem:EasyDirection} requires more technical detail.
We therefore introduce some further notation that will be used throughout the remainder of the section.
Assume our constructed instance of $1|B,r_j|\sum w_jU_j$ admits a \emph{solution schedule}, i.e., a schedule where the total weight of tardy jobs is at most $kX+(n-1)k-t$.
Let $B_1,\ldots,B_b,B_{b+1}$ denote the batches of this schedule, with respective starting times $S_1 < \cdots < S_{b+1}$ and completion times $C_1 < \cdots < C_{b+1}$.
Below we modify this schedule, without increasing the total weight of tardy jobs, in order to make our arguments easier.   
\begin{lemma}
\label{lem:normalized}%
  Suppose that the constructed instance of $1|B,r_j|\sum w_jU_j$ has a solution schedule.
  Then it has a solution schedule with batches $B_1,\ldots,B_b,B_{b+1}$, scheduled in that order, where:
  \begin{itemize}
    \item All tardy jobs are in $B_{b+1}$, and include no leftover jobs. 
    \item All early jobs are in $B_1,\ldots,B_b$, and include normal jobs with total weight at least $t+k$.
%\item For each $\ell \in \{1,\ldots,k\}$ with $E_\ell \neq \emptyset$, there is a unique batch $B \in \{B_1,\ldots,B_b\}$ with $E_\ell \subseteq B$.
  \end{itemize}
\end{lemma}
\iflong
\begin{proof}
  Consider any solution schedule with batches $B_1,\ldots,B_b$, scheduled in that order, that has at most $k X-t +(n-1)k$ total weight of tardy jobs.
  We first observe that no leftover job is tardy, as a single leftover job has weight $k(X+n) > k X-t +(n-1)k$.
  Moreover, as the total weight of all normal jobs of the instance is $k(X+n)$, the total weight of the early normal jobs must be at least $t+k$.
  Finally, we can move all tardy jobs to a new batch $B_{b+1}$ that starts right after $B_b$ completes, deleting all empty batches resulting from this, without increasing the total weight of tardy jobs. 
\end{proof}
\fi

Due to Lemma~\ref{lem:normalized}, some normal jobs must be early.
For $\ell \in \{1,\ldots,k\}$, we use $E_\ell$ denote the early jobs of type $\ell$ in the schedule.
Then $\bigcup_\ell E_\ell \neq \emptyset$.
We use $p(E_\ell)$ and $w(E_\ell)$ to respectively denote the total processing time and weight of jobs in $E_\ell$, i.e., $p(E_\ell)=\sum_{j \in E_\ell} p_j$ and $w(E_\ell)=\sum_{j \in E_\ell} w_j$.

\begin{lemma}
\label{lem:normalized2}%
  For each $\ell \in \{1,\ldots,k\}$ with $E_\ell \neq \emptyset$, there is a unique batch $B(\ell) \in \{B_1,\ldots,B_b\}$ with $E_\ell \subseteq B$.
\end{lemma}
\iflong
\begin{proof}
  Choose some non-empty $E_\ell$.
  Then each job $j \in E_\ell$ is released at time $r_\ell$ and has a due date of $r_\ell+t+x < r_\ell + 2t$ for some $x \in \{x_1,\ldots,x_n\}$ (the inequality follows as all $x_i$ are smaller than $t$).
  As batch setup requires $t$ time, and all jobs in $E_\ell$ are early, there must be some batch that contains all jobs of $E_\ell$.
  Furthermore, this batch cannot contain jobs of some $E_{\ell'}$, $\ell' \neq \ell$, since those jobs either have deadlines prior to $r_i$ (in case $\ell' < \ell$), or release times that are later than the due dates of jobs in $E_\ell$ (in case $\ell' > \ell$). 
\end{proof}
\fi

Lemma~\ref{lem:normalized2} implies that we can assume there is a specific batch associated with each non-empty $E_\ell$.
Let~$d_\ell$ be the earliest deadline in $E_\ell$.
Then $d_\ell = (\ell-1)3t+t+x_{\pi(\ell)}$ for some integer $x_{\pi(\ell)} \in \{x_1,\ldots,x_n\}$.
Thus, there is also a specific due date and input integer associated with~$E_\ell$. 
\begin{lemma}
  For each $\ell \in \{1,\ldots,k\}$ with $E_\ell \neq \emptyset$ we have: 
  \begin{itemize}
    \item $p(E_\ell) \leq x_{\pi(\ell)}$.
    \item $w(E_\ell) \leq p(E_\ell)+1$, and this holds with equality if and only if $E_\ell=J_{\pi(\ell), \ell}$.
  \end{itemize}
\end{lemma}
\iflong
\begin{proof}
  According to Lemma~\ref{lem:normalized2}, there is a unique batch $B(\ell)$ which includes all jobs of $E_\ell$.
  As the release time of all jobs in $E_\ell$ is $r_\ell=(\ell-1)\cdot 3t$, and the setup time of $B(\ell)$ is $t$, it must be that $p(E_\ell) \leq ||B(\ell)|| \leq x_{\pi(\ell)}$; otherwise, jobs in $E_\ell$ with due date $d_\ell$ would be late.
  Now, for each job $j \in E_\ell$, let $x(j) \in \{x_1,\ldots,x_n\}$ denote the integer for which $j$ is associated with (i.e., $j \in J_{\ell,x(j)}$).
  Then $x(j) \geq x_{\pi(\ell)}$ by definition of $x_{\pi(\ell)}$.
  Since $p(E_\ell) \leq x_{\pi(\ell)}$, we have
  \begin{equation*}
     w(E_\ell)    = \sum_{j \in E_\ell} p_j + p_j/x(j)
               \leq \sum_{j \in E_\ell} p_j + p_j/x_{\pi(\ell)}
                  = p(E_\ell) + p(E_\ell)/x_{\pi(\ell)} \leq p(E_\ell) + 1\enspace .
  \end{equation*}
  Note that the first inequality is strict if and only if there is a job $j\in E_\ell\setminus J_{\pi(\ell), \ell}$ as $x(j) \geq x_{\pi(\ell)}$ and the second inequality is strict if and only if $p(E_\ell) < x_{\pi(\ell)}$.
  Hence equality holds if and only if $E_\ell=J_{\pi(\ell), \ell}$.
  The statement of the lemma thus follows.
\end{proof}
\fi

\begin{lemma}
\label{lem:normalized3}%
  $E_\ell \neq \emptyset$ for each $\ell \in \{1,\ldots,k\}$.
\end{lemma}
\iflong
\begin{proof}
  By Lemma~\ref{lem:normalized}, we have $t+k \leq \sum_\ell w(E_\ell)$.
  By Lemma~\ref{lem:normalized2}, we have $p(E_\ell) \leq x_{\pi(\ell)}$ and $w(E_\ell) \leq p(E_\ell) + 1$.
  Thus, 
  \begin{equation*}
    t \leq \sum_{\ell=1}^k w(E_\ell) - k
      \leq \sum_{\ell=1}^k p(E_\ell)
      \leq \sum_{\ell=1}^k x_{\pi(\ell)},
  \end{equation*}
  where $x_{\pi(\ell)} = 0$ if $E_\ell = \emptyset$ in the summation above.
  Since any $k-1$ integers in $\{x_1,\ldots,x_n\}$ sum up to a number which is smaller than $t$, it must be that $x_{\pi(\ell)} > 0$ for all $\ell \in \{1,\ldots,k\}$, and the statement of the lemma follows.
\end{proof}
\fi

\begin{lemma}
\label{lem:normalized4}%
  If there is a solution schedule, then there is one with batches $B_1,\ldots,B_{2k+1}$ scheduled in that order, where for each $\ell \in \{1,\ldots,k\}$:
  \begin{itemize}
    \item $B_{2\ell-1}$ is scheduled at time $3t(\ell-1)$, and $B_{2\ell}$ is scheduled immediately after the completion of $B_{2\ell-1}$.
    \item $B_{2\ell-1}$ contains only normal jobs of type of $\ell$, and $B_{2\ell-1}$ contains only leftover jobs.
    \item All tardy jobs are in $B_{2k+1}$, and are normal.
  \end{itemize}
\end{lemma}
\iflong
\begin{proof}
  Let $B_1,\ldots,B_{b+1}$ be the batches of our schedule as in Lemma~\ref{lem:normalized}.
  We modify the batches of this schedule so as to fit the requirements of the lemma without increasing the total weight of tardy jobs in the schedule. 
  
  We first note that for each $\ell$ batch $B(\ell)$ is completely processed in the interval $[3t(\ell-1),\linebreak 3t(\ell-1)+2t]$.
  Thus, if there is no batch between $B(\ell)$ and $B(\ell+1)$, we might as well add one as the time between the completion time of $B(\ell)$ and the starting time of $B(\ell+1)$ is at least $t$.
  (We set $B(k+1)=B_{b+1}$.)
  Since there is a batch $B(\ell)$ for each $\ell \leq k+1$, by Lemma~\ref{lem:normalized3}, there are $2k$ batches consisting only of early jobs.

  Suppose that the completion time of a batch $B_i$ is in the interval $(3t(\ell-1), 3t(\ell-1)+t]$ for some $\ell$.
  Then $B_i$ cannot contain type $\ell$ jobs, as it started before $3t(\ell-1)$.
  Hence, $B_i$ only contains leftover jobs.
  We can move some leftover jobs from $B_i$ to $B_{i+1}=B(\ell)$ and simultaneously reduce the starting time of $B(\ell)$ by the number of moved jobs, until the starting time of $B_i$ equals $3t(\ell-1)$.

  If no batch is completed in $(3t(\ell-1), 3t(\ell-1)+t]$, then we can start $B(\ell)$ at time $3t(\ell-1)$.
  This can only decrease the completion times of the jobs.
  If there are leftover jobs in batch $B(\ell)=B_{2\ell-1}$, then we can move them to batch $B_{2\ell}$.
  They will not be late as the completion time of $B_{2\ell}$ is at most $3kt$.
%
%  We begin with the first batch $B_1$.
%  We first set $S_1 = 0$; clearly this does not make $B_1$ complete later. Now, by Lemma~\ref{lem:normalized3} we know that $E_1 \neq \emptyset$, and $E_1 \subseteq B(1)$ for some batch $B(1)$ by Lemma~\ref{lem:normalized2}.
%  Moreover, $B(1)$ cannot be processed after $B_1$, since otherwise it would complete later than time $2T > d_1$.
%  Thus, $B_1=B(1)$, and so $E_1 \subseteq B_1$.
%  We now remove all leftover jobs in $B_1$ according to one of the following two cases:
%  \begin{itemize}
%    \item $S_2 < r_2$: In this case, we add all leftover jobs of $B_1$ to $B_2$, and set $S_2=C_1$.
%    Note that $B_2$ does not complete later in this way. 
%    \item $S_2 > r_2$: In this case we create a new batch $B_2$ that starts at $C_1$ and contains all leftover jobs in $B_2$. 
%  \end{itemize}
%  Note that in the second case above, the new batch completes before the next batch starts, as
%  \begin{equation*}
%    2T + ||B_1|| \leq x_12T + p(E_1)+x_1 \leq 2(T+x_1) < 3T = r_2, 
%  \end{equation*}
%  due to ... \textcolor{red}{complete argument}
%  Thus, in both cases, the resulting schedule is feasible, and the number of tardy jobs %does not increase. 
%
%  Now let $\ell > 1$, and assume the lemma holds for $\ell' < \ell$.
%  By the same argument, we know that $B(\ell)=B_{2\ell}$.
%  It could be that \textcolor{red}{complete proof}
%  We proceed almost in the same thing.
%  The only difference
%  Let
\end{proof}
\fi

\begin{lemma}
\label{lem:HardDirection}%
  Suppose that the constructed instance of $1|B,r_j|\sum w_jU_j$ admits a schedule with $\sum_j w_jU_j \leq k X-t +(n-1)k$.
  Then there exist $x_{\pi(1)},\ldots, x_{\pi(k)} \in \{x_1,\ldots,x_n\}$ so that \mbox{$\sum_i x_{\pi(i)} = t$}. 
\end{lemma}
\iflong
\begin{proof}
  Let $B_1,\ldots,B_{2k+1}$ be the batches of a schedule as promised by Lemma~\ref{lem:normalized4} for our $1|B,r_j|\sum w_jU_j$ instance with $\sum_j w_jU_j \leq k X-t +(n-1)k$.
  Then batch $B_{2k}$ completes at time $C_{2k} \leq 3kt$, since $3kt$ is the latest due date of the input jobs.
  Since there are $2k$ batches with early jobs, and the setup time for each of these batches is $t$, we have $\sum_{\ell=1}^{2k} ||B_\ell|| \leq kt$.
  Thus, as the total processing times of all leftover jobs is $(k-1)t$, we have 
  \begin{equation*}
    \sum_{\ell=1}^k p(E_\ell) = \sum_{\ell=1}^{k} ||B_{2\ell-1}|| - \sum_{\ell=1}^{k} ||B_{2\ell}||
         =  \sum_{\ell=1}^{2k} ||B_\ell|| - (k-1)t
      \leq t \enspace .
  \end{equation*}
  Recall that, by Lemma~\ref{lem:normalized} and Lemma~\ref{lem:normalized2}, we also have
  \begin{equation*}
    t \leq \sum_{\ell=1}^k w(E_\ell) - k
      \leq \sum_{\ell=1}^k p(E_\ell) \enspace .
  \end{equation*}
  It follows that $\sum p(E_\ell) = t$, and $\sum w(E_\ell) = \sum p(E_\ell) + k$.
  The latter equality can only happen if $w(E_\ell) = p(E_\ell)+1$ for each $\ell \in \{1,\ldots,k\}$, which in turn implies by Lemma~\ref{lem:normalized3} that $p(E_\ell)=x_{\pi(\ell)}$ for each $\ell \in \{1,\ldots,k\}$.
  Thus, $\sum x_{\pi(\ell)} = t$, and the statement of the lemma follows. 
\end{proof}
\fi

\subsection{Parameter $\mathbf{\#p+\#d}$}
Lemma~\ref{lem:EasyDirection} and Lemma~\ref{lem:HardDirection} combined prove that our construction indeed shows $\mathsf{W}[1]$-hardness for parameter $\#p+\#r$.
We next show that this construction can be transformed to show hardness for parameter $\#p+\#d$.
\begin{lemma}
\label{relduelemma}
  For non-negative integers $k,k'$, any instance of $1|B, r_j|\sum w_jU_j$ with $k$ distinct release dates and $k'$ distinct due dates can be transformed into an instance of $1|B, r_j|\sum w_jU_j$ with $k'$ distinct release dates and $k$ distinct due dates, which has the same objective value.
\end{lemma}
\iflong
\begin{proof}
  Let $J$ be a set of $n$ jobs forming an instance of $1|B, r_j|\sum w_jU_j$.
  We create a set $J'$ of $n$ jobs, as follows.
  For each job $j\in J$ we create one job $j'\in J'$ with $p_{j'}=p_j$, $w_{j'}=w_j$, $r_{j'}=-d_j$ and $d_{j'}=-r_j$.
  Observe that the problem of finding a maximum-weight set of early jobs is the same for both $J$ and $J'$:

  Let $\sigma$ be a schedule for $J$, and let $J_e(\sigma)$ be the set of jobs in $J$ that are early in $\sigma$.
  For $j \in J_e(\sigma)$ let $S_j$ denote its starting time of $j$ and $C_j$ its completion time.
  Then we obtain a schedule $\sigma'$ for $J'$ by setting the start time of $j'$ to be $S_{j'}=-C_j$ for all jobs $j\in J_e(\sigma)$ and scheduling the remaining jobs late.
  No two jobs will be processed at the same time, as the intervals $(S_j, C_j),(S_{j'}, C_{j'})$ are pairwise disjoint for all $j,j'\in J_e(\sigma)$.
  Thus the intervals $(-C_j, S_j),(-C_{j'},S_{j'})$ are also pairwise disjoint for all $j,j'\in J'_e(\sigma')$.
  Further, for each $j \in J_e(\sigma)$ we have $S_j \geq r_j$ and $d_j \geq C_j$ and thus also $r_{j'}=-d_j\leq -C_j =S_{j'} $ and $d_{j'}=-r_j \geq -S_j=C_{j'}$.
  
  Similarly, given the set $J'_e(\sigma')$ of early jobs for a schedule $\sigma'$ for $J'$ we obtain a schedule for~$J$ such that all jobs $j$ for which $j' \in J'_e(\sigma')$ are scheduled early, by setting $S_j=-C_{j'}$.

  This shows that the problem $1|B, r_j|\sum w_jU_j$ with parameter $\#d + \#p$ is as hard as\iflong\linebreak\fi $1|B, r_j|\sum w_jU_j$ with parameter $\#r + \#p$.
\end{proof}
\fi

\begin{corollary}
  Problem $1|B, r_j|\sum w_jU_j$ is $\mathsf{W}[1]$-hard for parameter $\#d + \#p$.
\end{corollary}

\subsection{XP algorithms} 
Last in this section we give an $\mathsf{XP}$-algorithm for the problem $1|B, r_j|\sum w_jU_j$ parameterized by $\#p+ \#r+ \#w$.
We use the following notation:
Similarly to the due dates, we order the release dates as follows: $r^{(1)}<r^{(2)} \dots < r^{(\#d)}$.

\begin{lemma}
  Problem $1|B, r_j|\sum w_jU_j$ is solvable in time $n^{f(\#p, \#r, \#w)}$.
\end{lemma}
\iflong
\begin{proof}
  Let $I$ be the set of job types with respect to processing time, weight and release date.
  Let $U=\{v \in \{1, \dots, n\}^{I} \}$ and $V=\{v=(v_1, \dots, v_{\#r})\in U^{\#r}\mid\sum_{\ell=1}^{\#r}(v_\ell)_i\leq n_i\} $ denote the space of possible solution vectors.
  For each element $v\in V$ we decide whether it is possible to get a schedule that starts $(v_\ell)_i$ early jobs of type $i \in I$ in the interval $[r^{(\ell)},r^{(\ell+1)})$.
  
  First, notice that if such a schedule exists then we might assume that the jobs of types $i$ are scheduled in order of their due date and that only the $\sum_{\ell=1}^{\#r}(v_\ell)_i$ jobs of type $i$ with the latest due dates are scheduled early.
  Thus we know which jobs are started in each interval~$[r^{(\ell)},r^{(\ell+1)})$.

  Second, notice that if we schedule the jobs that are started in $[r^{(\ell)},r^{(\ell+1)})$ in (EDD)-order starting new batches only if it is necessary then we also get a schedule for these jobs that ends as early as possible.
  Thus all we need to do in order to decide whether such a schedule exists is to the following:
  First schedule all jobs that start in $[r^{(1)},r^{(2)})$ in (EDD).
  Then let $t_1$ be the date where the last of these jobs is finished.
  Then we schedule all jobs that start in $[r^{(2)},r^{(3)})$ in (EDD) but the starting time of the first batch is $\min \{r_2, t_1\}$. Then let $t_2$ be the date where the last of these jobs is finished.
  We then continue in the obvious way.
  If all jobs scheduled are early and no job is started before its release time then there is such a schedule; otherwise, no such schedule exists.
  From all schedules we obtain, we take the one that maximizes $\sum_{\ell, i}(v_\ell)_i w_i$.
  The total run time is $n^{O(\#r^2 \#p \#w)}$.
\end{proof}
\fi

Using Lemma \ref{relduelemma} we also get the following result:
\begin{corollary}
  Problem $1|B, r_j|\sum w_jU_j$ is solvable in $n^{f(\#p, \#d, \#w)}$ time. 
\end{corollary}

%%%%%%%%%%%%%%%%%%%%%%%%%%%%%%%%%%%%%%%%%%%%%%%%%%
%%%%%%% Section: Batch restrictions 
%%%%%%%%%%%%%%%%%%%%%%%%%%%%%%%%%%%%%%%%%%%%%%%%%%
\section{Batch restrictions}
\label{sec:batchrestrictions}
In this section we consider the variants of $1|B|\sum w_jU_j$ where the batches are either restricted in terms of their size ($|B|\leq b)$ or their volume ($||B||\leq b$).

In this section we will use the notion of job types: Each job $j\in J$ has a \emph{type}, which is given by the vector $\tau(j) = (p_j, w_j, d_j, r_j)$.
In some settings parts of the tuple can be omitted, which allows us to shortcut the job type.
For example, a job of type $(p_j, w_j, d_j, r_j)$ is also of type $(p_j, d_j, r_j)$.
We denote the set of all job types by $\mathcal{T}$.
%Throughout the paper, we assume a succinct input encoding for problem $1|B,r_j|w_jU_j$, which means that the batch setup time $\Delta$ is encoded in binary and the job set $J$ is encoded by a vector $\vec{n} \in \mathbb{N}^{\mathcal{T}}$ indicating the number of jobs of each type and a multi-dimensional vector $\vec{c}$ of job characteristics for each way.
For each type $\tau\in\mathcal T$ let $d_{\tau},p_{\tau},r_{\tau}$ and $w_{\tau}$ denote the due date, processing time, release date and weight of jobs with type $\tau$.
Note that if all jobs are released at time zero, then a schedule can be given by a function $\sigma:\{1, \dots \#d\} \rightarrow \mathbb{N}^{\mathcal{T}}$; let $\sigma(\ell)_\tau$ indicate the number of jobs of type $\tau$ that are completed in the time interval $(d^{(\ell-1)}, d^{(\ell)}]$.

\subsection{Bounded batch sizes}
First, we show that $1||B|\leq b|\sum w_jU_j$ is fixed-parameter tractable for parameter $ \#d+\#w$, proving the first part of Theorem\ref{thm:weighted-tardy-boundedbatchsize-dw}.
\begin{lemma}
  Problem $1||B|\leq b|\sum w_jU_j$ can be solved in time $f(\#d+\#w)\cdot n^{O(1)}$.
\end{lemma}
\iflong
\begin{proof}
  Given an instance $\mathcal{I}$ of $1|B,|B| \leq b|\sum w_jU_j$, we set up the following mixed-integer linear program (MILP) to find an optimal schedule.
  The variables of the MILP are defined as follows.
  Let $I=\{ (w, d) ~|~ (w, p, d)\in \mathcal{T} \text{ for some $d$} \}$ be the set of job types with respect to weight and due date. 
  For each type $i\in I$ and each $\ell\in\{1,\dots,\#d\}$ we have one integer variable $x_{i,\ell}$, indicating the number of jobs of type $i$ finishing job in the time interval $(d^{(\ell-1)}, d^{(\ell)}]$.
  (Note that this means that their batches finish in the interval.)
  For each job type $\tau = (d_\tau,p_\tau, w_\tau) \in \mathcal{T}$, we have one fractional variable $y_{(\tau,\ell)}\in [0, n_\tau]$ to indicate the number jobs of type $\tau$ which are processed in time before their due date $ d^{(\ell)}$.
  (Recall that $n_\tau$ is the number of jobs of type $\tau$.) 
  For each index $\ell\in \{1,\dots,\#d\}$ we have one integer variable $z_{\ell}$ to indicate the number of batches that are completed before or at time $ d^{(\ell)}$.
  Finally, we set $z_0=0$.
  
  The MILP is given by
  \begin{align}
                & \min \sum_{\tau \in \mathcal{T}} (n_\tau - y_{(\tau,\#d)})w_\tau \\
    \label{eq2} & z_{\ell}\geq z_{\ell-1}+\frac{1}{b} \sum_{i \in I} x_{i,\ell}, &  & \ell = 1,\dots,\#d,\\
    \label{eq3} & \sum_{\ell_0 \leq \ell}x_{i,\ell_0} = \sum_{\tau \in \mathcal{T}, p_\tau = p_i \wedge d_\tau=d_i}y_{(\tau,\ell)}, &   & i\in I, \ell = 1,\dots,\#d,\\
    \label{eq4} & z_{\ell}\Delta+\sum_{\tau \in \mathcal{T}}p_\tau y_{(\tau,\ell)}\leq d^{(\ell)} &  &\ell = 1,\dots,\#d \enspace .
  \end{align}
  The MILP has $\#d  (|I|+1) = O(\#d^2\cdot \#w)$ integer variables and $|\mathcal{T}| \#d = O(|\mathcal{T}|^2)$ fractional variables.
  It can be solved by Lenstra's algorithm~\cite{Lenstra1983} for integer programming in fixed dimension in time $f(\#d, \#w)\cdot n^{O(|\mathcal{T}|)}$.

  It remains to show that optimal solutions of value $W$ to the MILP correspond to optimal schedules with weighted number of tardy jobs equal to $W$.
  A crucial observation is that, given an optimal solution to the MILP, we can assume that all variables $y_{(\tau,\ell)}$ take integer values.
  This is due to the fact that, given a job type $\tau \in \mathcal{T}$ and an index $\ell \in \{1,\dots,\#d\}$, we can assume that if $y_{\tau\ell}< n_j$ then $y_{\tau'\ell}= 0$ for all $\tau'$ with $p_{\tau'}> p_\tau, w_{\tau'}= w_\tau$ and $d_{\tau'}= d_\tau$.
  For if that was not the case, then we can increase $y_{\tau\ell}$ and decrease $y_{\tau'\ell}$ by the same amount, without changing the objective value or violating constraint~\eqref{eq3} or constraint~\eqref{eq4}.
  The intuition here is that we can process the jobs of type $i \in I$ in increasing order of their processing time.

  Note that~\eqref{eq2} assures that we use $\lceil \frac{1}{b} \sum_{i \in I} x_{i\ell} \rceil$ batches ending in $(d^{(\ell-1)},d^{(\ell)}]$ which is the minimum number of batches needed to complete all the jobs ending in that interval.
  Constraint~\eqref{eq3} is for determining the exact types of the jobs that are processed rather than just the type with respect to weight and due date.
  As mentioned we can assume that the $y$-variables are integral in an optimum solution.
  Constraint~\eqref{eq4} makes sure that all the early jobs are indeed completed before their due date.

  To obtain a schedule from a solution to the MILP, we first process $x_{i,1}$ jobs of type $i$ for each $i$ in order of their processing times with ties broken arbitrarily, and always starting a new batch when necessary and closing the last batch at the end.
  Then we can continue with $x_{i,2}$ jobs of type $i$ for each $i$ the same way, and so on.
  Conversely, a schedule translates into a solution (also fulfilling~\eqref{eq2}) using the interpretations for the variables.
\end{proof}
\fi

If $\#p + \#d$ (rather than $\#w + \#d$) is bounded by our parameter then we get an even stronger result. More precisely we can solve instances where jobs additionally can have different release dates as long as the number of different release dates is also bound by our parameter.
\begin{theorem}
\label{thm:boundedbatchsize-reltimes-pr}
  Problem $1|B, |B| \leq b, r_j|\sum w_jU_j$ is fixed-parameter tractable for parameter\iflong\linebreak\fi $\#d + \#p + \#r$.
\end{theorem}
\iflong
\begin{proof}
  We set $T=\{ r_j~|~ j \in J \}\cup\{ d_j~|~ j \in J \}$ to be the set of critical time points.
  Further we order $T=\{t_1, \dots , t_k\}$ in increasing order, i.e., $t_1 < t_2 < \dots < t_k$.
  We again design a MILP, but this time with slightly more variables.
  Instead of variables $z_\ell$, this time we will use integral variables~$z_{\ell, \ell'}$ for any $\ell<\ell'$ to indicate the number of batches that start at or after $t_{\ell}$ but before $t_{\ell+1}$ and finish before or at $t_{\ell'}$ but after~$t_{\ell'-1}$.  
  Now we set $I=\{ (p, r, d) ~|~ (p, w, r, d)\in \mathcal{T} \text{ for some $d$} \}$ be the set of job types with respect to weight and due date.
  Instead of $x_{i, \ell}$, we have integral variables $x_{i, \ell, \ell'}$ to indicate the number of early jobs of a given type that are processed in batches starting at or after $t_{\ell}$ but before $t_{\ell+1}$ and completed before or at $t_{\ell'}$ but after $t_{\ell'-1}$.
  We note that we remove variables $x_{i, \ell, \ell'}$ if $d_i<t_{\ell'}$ or $r_i>t_\ell$.
  We use variables $y_\tau$ to indicate the number of early jobs of type $\tau\in \mathcal{T}$. 
  %For release date $r$ we denote by $d(r)$ the smallest due date greater than~$r$.
  
  The MILP has the following constraints: 
  \begin{align*}
    & \min \sum_{\tau \in \mathcal{T}} (n_\tau - y_{\tau})w_\tau \\
    & \sum_{i \in I}x_{i, r, d} \leq b z_{\ell, \ell'}                                                      &  & \text{for any $1 \leq \ell< \ell' \leq k$} \\ 
    & \sum_{\ell, \ell'} x_{i, \ell, \ell'} = \sum_{\tau \in \mathcal{T}, w_\tau=w_i \wedge r_i=r_\tau \wedge d_\tau=d_i }y_{\tau} &  & \text{for each $i \in I$} \\ 
    & t+\sum_{t\leq t_\ell < t_{\ell'} \leq t'}\left(z_{\ell, \ell'}\Delta+\sum_{i\in I}p_ix_{i,\ell ,\ell'}\right)\leq t' &  & \text{for each $t, t' \in T$ with $ t < t'$}\\
    & y_{\tau}  \leq n_\tau &  & \text{for each $\tau\in \mathcal{T}$} 
  \end{align*}
  We further need two more kinds of constraints to guarantee that if there is a long batch, i.e., a batch that starts before $t_{\ell}$ and ends at or before $t_{\ell'}$ but after $t_{\ell'-1}\geq t_{\ell}$, then there cannot be any other batch starting and ending in $[t_{j}, t_{j'}]$ for any pair $(j, j') \in \{\ell, \dots, \ell'\}^2\setminus \{ (t_{\ell}, t_{\ell+1}), (t_{\ell'-1}, t_{\ell'}) \}$.
  \begin{align}
  \label{eq10}
    & z_{\ell_1, \ell_2}+z_{\ell_3, \ell_4} \leq 1  &  & \text{if $\ell_1 \leq \ell_3 <\ell_3+2 \leq \ell_4 \leq \ell_2 $}\\ \label{eq11}
    & \frac{z_{\ell, \ell+1}}{n}+z_{\ell_1, \ell_2} \leq 1  &  & \text{if $\ell_1 < \ell $ and $\ell_2>\ell+1$} \enspace .
  \end{align}
  
  Using the interpretations of the variables given a schedule, one can easily construct a feasible solution of the MILP with same value.
  
  We claim that in any optimal solution of the MILP, all variables of the form $y_{\tau}$ are integral, and $y_\tau \leq n_\tau$ implies $y_{\tau'}=0$ for all other types $\tau'$ with the same processing time, release date and due date but higher weight.
  For proof, suppose, for sake of contradiction, that there is some non-integral~$y_\tau$.
  Let $\tau$ be of (sub)type $i \in I$.
  Since $\sum_{\tau \in \mathcal{T}, w_\tau=w_i \wedge r_i=r_\tau \wedge d_\tau=d_i }y_{\tau} =\sum_{\ell, \ell'} x_{i, \ell, \ell'}$ is integral there must be another non integral variable $y_{\tau'}$ such that $\tau'$ is also of type $i$.
  Assume, without loss of generality, that $w_\tau>w_{\tau'}$.
  Now since $n_\tau$ is integral, we have $y_\tau < n_\tau$.
  Thus we can increase~$y_\tau$ and decrease $y_{\tau'}$ by the same amount until either $y_\tau=n_\tau$ or $y_{\tau'}=0$.
  The solution we get is still feasible, but its value is smaller, contradicting the optimality of the initial solution.
  The same argumentation can be used to show that $y_\tau \leq n_\tau$ implies $y_{\tau'}=0$ for all other types $\tau'$ with the same processing time, release date and due date but higher weight.
  This proves the claim.

  Now to create a schedule we create $z_{\ell, \ell'}$ batches $\mathcal{B}_{\ell, \ell'}$ for each variable $z_{\ell, \ell'}$ and fill them with appropriate jobs, i.e., such that there $x_{i, \ell, \ell'}$ jobs of type $i$ assigned to them.
  We schedule the batches in the following way: If batch $B$ is in $\mathcal{B}_{\ell, \ell'}$ and batch $B'$ is in $\mathcal{B}_{\ell_1, \ell_2}$ then we schedule~$B$ before $B'$ if $\ell < \ell_1$, or $\ell=\ell_1$ and $\ell'<\ell_2$, breaking ties arbitrarily.
  Given this ordering, we schedule batch $B \in \mathcal{B}_{\ell, \ell'}$ at the completion time of the previous batch if it finishes later than $t_\ell$, or at time $t_\ell$ otherwise.

  We need to show that indeed all $\sum_{\ell, \ell'} x_{i, \ell, \ell'} $ jobs of type $i$ scheduled in these kind of batches are early for each type $i$.
  Suppose, for sake of contradiction, that there is late job $j$ in batch $B \in \mathcal{B}_{\ell, \ell'}$ for some $\ell$ and $\ell'$.
  Let $t_{\ell_0}$ be the latest time point less or equal to $t_\ell$ such that there is idle time before $t_{\ell_0}$, or---if no such time exists---we set $t_{\ell_0}$ to be the smallest release time.

  We claim that
  \begin{equation*}
    t_{\ell_0}+\sum_{{\ell_0}\leq \ell_1, \ell_2 \leq \ell'}\left(z_{\ell_1, \ell_2}\Delta+\sum_{i \in I}p_ix_{i,\ell_1 ,\ell_2}\right)> t_{\ell'} \enspace .
  \end{equation*}
  To see that notice, that only jobs in batches in $ \mathcal{B}_{\ell_1, \ell_2}$ with $\ell_1 \geq \ell_0$ and $\ell_2 \leq \ell'$ are scheduled before the completion time of $j$.
  This holds true as any batch $B' \in \mathcal{B}_{\ell_1, \ell_2}$ with $\ell_1<\ell_0$ is completed before $t_{\ell_0}$ by definition of $\ell_0$, and any batch $B' \in \mathcal{B}_{\ell_1, \ell_2}$ with $\ell_1\geq \ell$ and $\ell_2 > \ell'$ is scheduled later than $j$.
  Further, we have $z_{\ell_1, \ell_2}=0$ if $\ell_1<\ell$ and $\ell_2>\ell'$ by constraint \eqref{eq10} and \eqref{eq11} using that $z_{\ell, \ell'}\geq 1$.
  However, our claim contradicts the feasibility of our solution thus $j$ cannot be late.
\end{proof}
\fi

\subsection{Bounded batch volume}
In the last part of this section we show why the problem $1|B |\sum U_j$ becomes hard when we add a bound to the maximum batch volume.
First, we consider parameter $\#d+\#w$, and afterwards parameter~$\#p$.

In {\sc Partition} we are given a set $T=\{x_1, \dots x_n\}$ of natural numbers such that $\sum_{x \in T}x=2K$; the task is to decide if there exists a set $T'\subseteq T$ such that $\sum_{x \in T'} x = \sum_{x \in T \setminus T'} x=K$.

We now devise a reduction from {\sc Partition} to show hardness of batch scheduling even in the unweighted case and a single due date.

\iflong
  \clearpage
  \pagebreak
\fi

\begin{theorem}
  Problem $1|B, ||B|| \leq V|\sum U_j$ is $\mathsf{NP}$-hard for $\#d=1$.
  % and there is no approximation algorithm without an additive term.
\end{theorem}
\iflong
\begin{proof}
  Let $(T=\{x_1, \dots x_n\};K)$ be an instance of {\sc Partition}; we construct an instance of $1|B, ||B|| \leq V|\sum U_j$ as follows:
  \begin{itemize}
    \item We set $\Delta=1$ and $V=K+1$.
    \item For each number $x_i$ there is one job $j_i$ with $p_{j_i}=x_i$ and $d_{j_i}=2K+2$.
  \end{itemize}
  Observe that there is a schedule with zero tardy jobs if and only if there is a subset $T'\subseteq T$ such that $\sum_{x \in T} x = \sum_{x \in T' \setminus T} x=K$.
  This is due to the fact that the only way to get such a schedule is to use exactly two batches of volume $V$ and a batch can only be of volume $V$ if the processing times of the jobs assigned to it add up to $K$.
\end{proof}
\fi

For parameter $\#p$ we prove the following result:

\begin{theorem}
\label{thm:batch-volumepd}
  Any instance $\mathcal I$ of $P||C_{\max}$ with $\#p$ different processing times can be transformed to an instance of $1|B,||B|| \leq V|\sum U_j$ with $\#p$ different processing times and a single due date, such that all jobs of $\mathcal I$ complete by time $T$ if and only if all jobs of $\mathcal I'$ are early.
\end{theorem}
\iflong
\begin{proof}
  Consider an instance $\mathcal{I}$ of $P||C_{\max}$ with job set $J$, number $m$ of machines, and target makespan $T$.
  We create an instance $\mathcal{I}'$ of $1|B,||B|| \leq V|\sum U_j$ consisting of a batch setup time~$\Delta$, a batch volume $V$, and a job set $J'$.
  We set $\Delta=Tm$ and $V=T$.
  The set $J'$ contains one job $j'$ for each job $j\in J$, where the processing time of $j'$ is the same as the processing time of~$j$ and the due date of $j'$ is equal to $d_j = mV$.

  In the forward direction, any schedule for $\mathcal{I}$ with makespan at most $T$ can be translated to a feasible schedule for $\mathcal{I}'$ that schedules all jobs early by creating one batch $B$ for each machine~$i$.
  All jobs scheduled on~$i$ will be assigned to $B$.
  Then the batch volume of each batch is at most~$T$, and all $m$ batches are completed early.
  
  In the backward direction, any schedule for $\mathcal{I}'$ has at most $m$ batches with early jobs, as
  \begin{equation*}
    (m+1)\Delta = (m+1)Tm
           > m(Tm + T)
           = m(\Delta+T)
           = mV
           = d_j \enspace .
  \end{equation*}
  Thus, for each batch $B$ with early jobs we can schedule all jobs assigned to $B$ on one machine, whose completion time is at most $T$.
  In summary, for $m$ batches with early jobs, we obtain a schedule for $\mathcal{I}$ with makespan at most $T$.
\end{proof}
\fi

\section{Discussion and Open Problems}
\label{sec:discussion}
We provided an extensive multivariate analysis of the single-machine batch scheduling problem to minimize the weighted number of tardy jobs.
In particular, we significantly refined and extended the work of Hochbaum and Landy~\cite{HochbaumLandy1994}, as well as Hermelin et al~\cite{HermelinEtAl2018}.

Several open questions remain, even for the setting without batches.
It appears especially challenging to resolve the question of whether $1||w_jU_j$ is fixed-parameter tractable for $\#p$, or~$\#w$, or turns out to be $\mathsf{W}[1]$-hard for either of those parameterizations.
This question was already stated by Hermelin et al.~\cite{HermelinEtAl2018}, and is not resolved here.
Naturally, we do not know the answer to this question for the more general $1|B|\sum U_jw_j$ problem; however, we also do not know the status of parameter $\#p + \#w$ for which $1||\sum w_jU_j$ is known to be fixed-parameter tractable~\cite{HermelinEtAl2018}. 
%We note that our approach by reudcing to the non-batch case which Hermelin et al.~\cite{HermelinEtAl2018} show that $1||\sum U_jw_j$ is fixed-parameter tractable for $\#p+\#w$ does not generalize to $1|B|\sum U_jw_j$, as following lemma shows:
%\begin{lemma}
%\label{lem:batchwp}%
%There is an instance of $1|B|\sum U_j$ for which the set $X\subseteq\{(x_i)_{i=(p, w)} ~|~ \text{there is}$\linebreak $\text{a schedule such that there are $x_i$ early jobs of type $i$.}\}$ is not convex.
%Thus there is no general MILP for $1| B | \sum U_j$ that has integral variables for each item type $i$.
%\end{lemma}
%\iflong
%\begin{proof}
%Consider the instance with $\Delta=4$, $p_1=p_2=1$, $d_1=d_2=6$, $p_3=p_4=2$, $d_3=d_4=8$.
%Here there are two jobs types and the two best solutions are to schedule either both jobs with $p_j=1$ early or both jobs with $p_j=2$ early.
%However it is not possible to schedule one job of each type early which is convex combination of both solutions.
%\end{proof}
%\fi
Another interesting question is to see if $1||B| \leq b|\sum U_j$ is fixed-parameter tractable for parameter $\#p$ or~$b$, or even solvable in polynomial time.


\iflong
  \clearpage
  \pagebreak
  \bibliographystyle{abbrvnat}
\else
  \bibliographystyle{splncs04}
\fi
  \bibliography{references}

\end{document}